\documentclass{PoS}

\newcommand{\be}{\begin{eqnarray}}
\newcommand{\ee}{\end{eqnarray}}
\newcommand{\nee}{\nonumber\end{eqnarray}}
\newcommand{\nn}{\nonumber\\}

\newcommand{\drbar}{{\overline{\rm DR}}}

\newcommand{\mch}[1] {m_{\ti \x^+_{#1}}}
\newcommand{\mnt}[1] {m_{\ti \x^0_{#1}}}

\newcommand{\msg}    {m_{\ti g}}
\newcommand{\msu}[1] {m_{\ti u_{#1}}}
\newcommand{\msd}[1] {m_{\ti d_{#1}}}

\def\gev             {{\rm GeV}}

\def\be            {\begin{equation}}
\def\ee            {\end{equation}}
\def\bea            {\begin{eqnarray}}
\def\eea            {\end{eqnarray}}

\def\a              {\alpha}
\def\b               {\beta}
\def\d               {\delta}

\def\x               {\chi}
\def\ti              {\tilde}
\def\sq              {\ti q}

\def\barb              {\bar{b}}

\def\barc             {\bar{c}}

\def\dll            {\d^{LL}_{23}}
\def\durr            {\d^{uRR}_{23}}
\def\durl            {\d^{uRL}_{23}}
\def\dulr            {\d^{uLR}_{23}}

\newcommand{\AddrVienna}{
\it Universit\"at Wien, Fakult\"at f\"ur Physik,
A-1090 Vienna, Austria \\}
\newcommand{\AddrGAKUGEI}{%
 \it Department of Physics, Tokyo Gakugei University, Koganei,
Tokyo 184-8501, Japan\\}

\newcommand{\AddrHEPHY}{%
 \it Institut f\"ur Hochenergiephysik der \"Osterreichischen Akademie
der Wissenschaften, A-1050 Vienna, Austria\\}

\title{Quark-flavour violation in $h^0 \to b \bar{b}$ in the MSSM at one-loop level}

\ShortTitle{Quark-flavour violation in $h^0 \to b \bar{b}$ in the MSSM at one-loop level}

\author{\speaker{E. Ginina}, H. Eberl, W. Majerotto \\
       \AddrHEPHY 
       E-mail: \email{elena.ginina@oeaw.ac.at, helmut.eberl@oeaw.ac.at, walter.majerotto@oeaw.ac.at}}
\author{A. Bartl \\
        \AddrVienna
        E-mail: \email{alfred.bartl@univie.ac.at}}
\author{K. Hidaka \\
        \AddrGAKUGEI
        E-mail: \email{hidaka@u-gakugei.ac.jp}}

\abstract{We compute the width of the decay $h^0 (125 ~{\rm GeV}) \to b \barb$ at next-to-leading order in the general MSSM with quark-flavour violation (QFV). We study the effect of mixing between the second and the third generation of squarks, taking into account the constraints on QFV from B-meson data. We discuss the renormalisation of the process as well as the resummation of the bottom Yukawa coupling at large $\tan \b$. We show numerical results on the decay width $\Gamma(h^0 \to b \barb)$ as a function of the involved QFV parameters and compare them with the corresponding width in the Standard Model.}

\FullConference{The European Physical Society Conference on High Energy Physics\\
		 22-29 July 2015\\
		 Vienna, Austria}

\begin{document}

\section{Introduction}

At present, our knowledge about the fundamental particles and their interactions is very well described by the Standard model (SM)
of particle physics. The experimental proof of the model was completed by the recent discovery of the Higgs boson at the LHC experiments~\cite{HiggsAtlas, HiggsCMS}. 
However, it is possible that the Higgs boson properties to be measured deviate from the SM, which would point to "New Physics".
Hence, it is important to determine if the observed Higgs boson is only part of the SM or of a more general theory.

The lightest neutral Higgs boson of the Minimal Supersymmetric Standard Model (MSSM), $h^0$, may very well coincide with the Higgs boson observed. Usually, its decays into quark-pairs are considered to be quark-flavour conserving (QFC) or at most minimally quark-flavour violating\footnote{Conceptually, all terms of minimal flavour-violation, both in SM and MSSM, are proportional to the CKM matrix.}. Yet the MSSM has an unexploited parameter potential to violate the quark-flavour symmetry in a non-minimal way. Generation mixing in the squark sector may significantly influence the decay widths of $h^0$ at one-loop level and can lead to quark-flavour violating (QFV) decays with sizeable rates. 
We have recently found that  $\Gamma(h^0 \to c \barc)$ can deviate by $\sim35\%$ from its SM rate due to QFV in the general MSSM~\cite{Bartl:2014bka
}. In this study we are looking for
 similar effects in the decay of $h^0$ into a pair of bottom quarks.

\section{Quark-flavour violation in the squark sector of the MSSM}
\label{sec:qfv}

In the super-CKM basis of MSSM, $\sq_{0 \gamma} =
(\sq_{1 {\rm L}}, \sq_{2 {\rm L}}, \sq_{3 {\rm L}}$,
$\sq_{1 {\rm R}}, \sq_{2 {\rm R}}, \sq_{3 {\rm R}}),~\gamma = 1,...6,$  
and $(q_1, q_2, q_3)=(u, c, t),$ $(d, s, b)$, the squark mass matrices have the following most general $3\times3$-block form
\bea
    {M}^2_{\tilde{u}} &=& \left( \begin{array}{cc}
        V_{\rm CKM} M_Q^2 V_{\rm CKM}^{\dag} +\cos 2\beta m_Z^2 (\frac{1}{2}-\frac{2}{3}
\sin^2\theta_W){\bf 1} + \hat{m}^2_u &\frac{v_2}{\sqrt{2}} T_U - \mu \hat{m}_u\cot\beta \nn \vspace{-4mm}\\ 
        \frac{v_2}{\sqrt{2}} T_U - \mu^* \hat{m}_u\cot\beta &M_U^2 + \frac{2}{3}\sin^2\theta_W \cos 2\beta m_Z^2 {\bf 1} + \hat{m}^2_u \end{array} \right), 
        \eea
       \bea  
    {M}^2_{\tilde{d}} &=& \left( \begin{array}{cc}
        M_Q^2 - \cos 2\beta m_Z^2 (\frac{1}{2}-\frac{1}{3}
\sin^2\theta_W){\bf 1} + \hat{m}^2_d &\frac{v_1}{\sqrt{2}} T_D - \mu \hat{m}_d\tan\beta \vspace{2mm}\\ 
       \frac{v_1}{\sqrt{2}} T_D - \mu^* \hat{m}_d\tan\beta &M_D^2 - \frac{1}{2} \sin^2\theta_W \cos 2\beta m_Z^2{\bf 1} + \hat{m}^2_d \end{array} \right). 
 \label{sqmass}       
\eea
The quark-flavour violating (off-diagonal) terms are contained in the soft SUSY-breaking mass matrices, $M_{Q, U, D}$, as well as in the soft SUSY-breaking trilinear coupling matrices $T_{U, D}$. The squark mass matrices (\ref{sqmass}) are diagonalized by the $6\times6$ unitary matrices $U^{\tilde{q}}$,
$\tilde{q}=\tilde{u},\tilde{d}$, such that
\begin{eqnarray}
&&U^{\tilde{q}} { M}^2_{\tilde{q}} (U^{\tilde{q} })^{\dag} = {\rm diag}(m_{\tilde{q}_1}^2,\dots,m_{\tilde{q}_6}^2)\,,
\end{eqnarray}
where $m_{\tilde{q}_1} < \dots < m_{\tilde{q}_6}$ are the physical masses of 
the flavour-mixed eigenstates $\sq_i=  U^{\sq}_{i \alpha} \sq_{0\alpha},~ i=1,...,6$.

For the quantitative treatment of the QFV effects
we introduce the following parameters in up-type (down-type) squark sector
\bea
\delta^{LL}_{\alpha\beta} & \equiv & M^2_{Q \alpha\beta} / \sqrt{M^2_{Q \alpha\alpha} M^2_{Q \beta\beta}}~,
\\[3mm]
\delta^{uRR}_{\alpha\beta} &\equiv& M^2_{U \alpha\beta} / \sqrt{M^2_{U \alpha\alpha} M^2_{U \beta\beta}}~,
\\[3mm]
\delta^{uRL}_{\alpha\beta} = \delta^{uLR*}_{\beta \a} &\equiv& (v_2/\sqrt{2} ) T_{U\alpha \beta} / \sqrt{M^2_{U \alpha\alpha} M^2_{Q \beta\beta}}~,
\\[3mm]
\delta^{dRR}_{\alpha\beta} &\equiv& M^2_{D \alpha\beta} / \sqrt{M^2_{D \alpha\alpha} M^2_{D \beta\beta}}~,
\\[3mm]
\delta^{dRL}_{\alpha\beta} = \delta^{dLR*}_{\beta \a}&\equiv& (v_1/\sqrt{2} ) T_{D\alpha \beta} / \sqrt{M^2_{D \alpha\alpha} M^2_{Q \beta\beta}}~.
\label{eq:dRL}
\eea
The subscripts $\alpha,\beta=1,2,3 ~(\alpha \ne \beta)$ denote the quark flavours $u,c,t$ ($d,s,b$) .

\section{Parameter constraints}
\label{sec:constr}

It is well known that in the SM flavour changing neutral current (FCNC) processes are strongly suppressed, imposing stringent constraints on quark generation-mixing. Clearly, any of the SM extensions must also obey these constraints. 
Although in the MSSM mixing between the first and the second generation of squarks is heavily constrained by the data on K-physics, there is still room for appreciable mixing between the second and the third squark generations consistent with the B-meson data. In this study we assume such a mixing, taking into account all constraints from B-physics experiments. We also take into account the LHC exclusion limits on the supersymmetric (SUSY) particle masses from direct searches as well as the experimental limit on SUSY contributions to the electroweak $\rho$ parameter~\cite{PDG}. Furthermore, we also respect the theoretical constraints from the vacuum stability conditions on the trilinear coupling matrices~\cite{Casas:1996de}.

\section{The process}

We study the decay of the lightest neutral Higgs boson, $h^0$, into a bottom-quark pair, $h^0 \to b \barb$, at full one-loop level in the general MSSM with quark-flavour violation. We calculate all SUSY-electroweak and SUSY-QCD one-loop corrections assuming mixing between the second and the third squark generations, namely $\ti{c}-\ti{t}$ and $\ti{s}-\ti{b}$ mixing. The decay width of $h^0$ is given by
\be
\Gamma(h^0 \to b \barb) = \Gamma^{\rm tree} + \d \Gamma^{\rm 1loop}\,,
\ee
where $\Gamma^{\rm tree}$ is the tree-level width and the flavour symmetry is broken by one-loop SUSY contributions to $\d \Gamma^{\rm 1loop}$ with up- and down-type squarks. QFV is generated by their mixing and is due to non-zero values of the off-diagonal elements of the matrices $M_{Q, U, D}$ (see Section~\ref{sec:qfv}). Moreover, the QFV parameters $T_{U\alpha \beta}$ and $T_{D\alpha \beta}$, for $\a \neq \b$, enter explicitly the interaction Lagrangian of $h^0$ with squarks.

\section{Renormalisation}

Loop calculations induce ultraviolet (UV) and infrared (IR) divergences. 
To deal with the UV divergences 
we adopt the $\drbar$ (dimensional reduction) renormalization scheme, which is most suitable for the MSSM.
The tree-level input parameters of the Lagrangian are defined at the scale $Q=1$~TeV. The shifts from the  $\overline{\rm DR}$-running parameters to the physical scale-independent parameters are obtained by imposing appropriate fixing conditions. Furthermore, to make our process free of IR divergences, we include the real gluon and photon radiation contributions by using a small gluon (photon) mass as a cutoff parameter.

\section{Resummation of the bottom Yukawa coupling at large $\tan \b$}

If the MSSM parameter $\tan \b$ is large the enhanced loop contributions need to be resummed~\cite{Carena:1999py}. To do the resummation we construct an effective Lagrangian, with the one-loop part given by
\be
{\cal L} = \frac{m_b}{\sqrt{2}v_1}\sin \a
\left(1+\frac{1}{\tan \a \tan \b}\right) \Delta_b \tan \b h^0 \barb b\,,
\label{efflag}
\ee
where $\Delta_b \tan \b = \Sigma_b^{LR}$ denotes the $\tan \b$-enhanced terms. For the resummed contribution we obtain
\be
\d \Gamma^{\rm > 1loop} = \Gamma^{\rm tree}  \left(\frac{1}{(1-\Delta_b \tan \b)^2} \left(1+\frac{\Delta_b}{\tan \a}\right)^2 - 2 \left(1+\frac{1}{\tan \a \tan \b}\right)\Delta_b \tan \b-1\right)\,,
\ee
where in $\d \Gamma^{\rm > 1loop}$ we have split away the one-loop part from the higher order corrections.
It is easy to check that $\d \Gamma^{\rm > 1loop}$ as well as the effective one-loop coupling given in (\ref{efflag}) vanish in the decoupling limit, $\tan \a \tan \b \to -1$, which is favoured by the measured Higgs boson mass.

\section{Total result}

Our total analytical result reads
\be
\Gamma(h^0 \to b \bar{b}) = \Gamma^{\rm tree}(m_b)+\d \widetilde{\Gamma}^g + \d \Gamma^{\tilde{g}} + \d \Gamma^{\rm EW}\,,
\label{fullres}
\ee
where $ \d \Gamma^{\tilde{g}}$ and $\d \Gamma^{\rm EW}$ denote the gluino and electroweak one-loop contributions, respectively, and $\d \widetilde{\Gamma}^g $ denotes the gluon contribution given in terms of $m_b\vert_{\rm SM}$
\be
\d \widetilde{\Gamma}^g =  \d \Gamma^{g} (m_b\vert_{\rm SM})-\Gamma^{\rm tree}(m_b) \frac{m_b^2-m_b^2\vert_{\rm SM}}{m_b^2}\,.
\ee
In $\d \widetilde{\Gamma}^g$ we have absorbed the large logarithm, $\ln (m_b/m_h^0)$, in the gluon one-loop contribution to the bottom mass, $\delta m_b^g$~\cite{Eberl:1999he}. Furthermore, we have checked that in $\Gamma^{\rm tree} + \d \Gamma^{\tilde{g}} $ the gluino mass dependence vanishes for $m_{\ti{g}} \to \infty$. The same holds for the  chargino and neutralino contributions as well.

\section{Numerical analysis}

In the following we show the output of one particular scenario, which we have found suitable to demonstrate the effect of QFV in $h^0 \to b \barb$. The scenario choice is consistent with the most recent Higgs boson mass measurement and all parameter constraints mentioned in Section~\ref{sec:constr} are satisfied. The MSSM input parameters at the scale $Q=1$~TeV are shown in Table~\ref{basicparam}. Table~\ref{physmasses} contains the physical output masses of the SUSY particles. %
The choice $Q =$ 1 TeV has been made in order to avoid instabilities caused by tadpole contributions in the calculation of $m_{A^0}^\drbar (Q)$ from the input parameter $m_{A^0}^{\rm pole}$.

\begin{table}[h!]
\begin{center}
\small{
\begin{tabular}{|c|c|c|}
  \hline
 $M_1$ & $M_2$ & $M_3$ \\
 \hline \hline
 600~\gev  &  1200~\gev & 2200~\gev \\
  \hline
\end{tabular}
\vskip0.4cm
\begin{tabular}{|c|c|c|}
  \hline
 $\mu$ & $\tan \beta$ & $m_{A^0}$ \\
 \hline \hline
  1500~\gev & 30  &    1500~\gev \\
  \hline
\end{tabular}
\vskip0.4cm
\begin{tabular}{|c|c|c|c|}
  \hline
   & $\alpha = 1$ & $\alpha= 2$ & $\alpha = 3$ \\
  \hline \hline
   $M_{Q \alpha \alpha}^2$ & ${3200}^2~\gev^2$ &  ${3000}^2~\gev^2$  & ${2200}^2~\gev^2$ \\
   \hline
   $M_{U \alpha \alpha}^2$ &  ${3200}^2~\gev^2$ &  ${3000}^2~\gev^2$  & ${2200}^2~\gev^2$ \\
   \hline
   $M_{D \alpha \alpha}^2$ &  ${3200}^2~\gev^2$ &  ${3000}^2~\gev^2$  & ${2500}^2~\gev^2$ \\
   \hline
\end{tabular}
\vskip0.4cm
\begin{tabular}{|c|c|c|c|}
  \hline
   $\delta^{LL}_{23}$ & $\delta^{uRR}_{23}$  &  $\delta^{uRL}_{23}$ & $\delta^{uLR}_{23}$\\
  \hline \hline
    0.17 & 0.8 & -0.025  &  -0.015 \\
    \hline
\end{tabular}
}
\end{center}
\caption{Input MSSM parameters 
at $Q = 1~{\rm TeV} $
with $T_{U33} = 1550$~GeV.}
\label{basicparam}
\end{table}
%
\begin{table}[h!]
\begin{center}
\small{
\begin{tabular}{|c|c|c|c|c|c|}
  \hline
  $\mnt{1}$ & $\mnt{2}$ & $\mnt{3}$ & $\mnt{4}$ & $\mch{1}$ & $\mch{2}$ \\
  \hline \hline
  $604$ & $1236$ & $1508$ & $1520$ & $1236$ & $1520$ \\
  \hline
\end{tabular}
\vskip 0.4cm
\begin{tabular}{|c|c|c|c|c|}
  \hline
  $m_{h^0}$ & $m_{H^0}$ & $m_{A^0}$ & $m_{H^+}$ \\
  \hline \hline
  $125$  & $1499$ & $1500$ & $1503$ \\
  \hline
\end{tabular}
\vskip 0.4cm
\begin{tabular}{|c|c|c|c|c|c|c|}
  \hline
  $\msg$ & $\msu{1}$ & $\msu{2}$ & $\msu{3}$ & $\msu{4}$ & $\msu{5}$ & $\msu{6}$ \\
  \hline \hline
  $2341$ & $1133$ & $2170$ & $3058$ & $3209$ & $3211$ & $3560$ \\
  \hline
\end{tabular}
\vskip 0.4cm
\begin{tabular}{|c|c|c|c|c|c|}
  \hline
 $\msd{1}$ & $\msd{2}$ & $\msd{3}$ & $\msd{4}$ & $\msd{5}$ & $\msd{6}$ \\
  \hline \hline
  $2158$ & $2515$ & $3008$ & $3059$ & $3207$ & $3212$ \\
  \hline
\end{tabular}
}
\end{center}
\caption{Physical mass spectrum in GeV.}
\label{physmasses}
\end{table}

\begin{figure*}[h!]
\centering
{ \mbox{\resizebox{7.4cm}{!}{\includegraphics{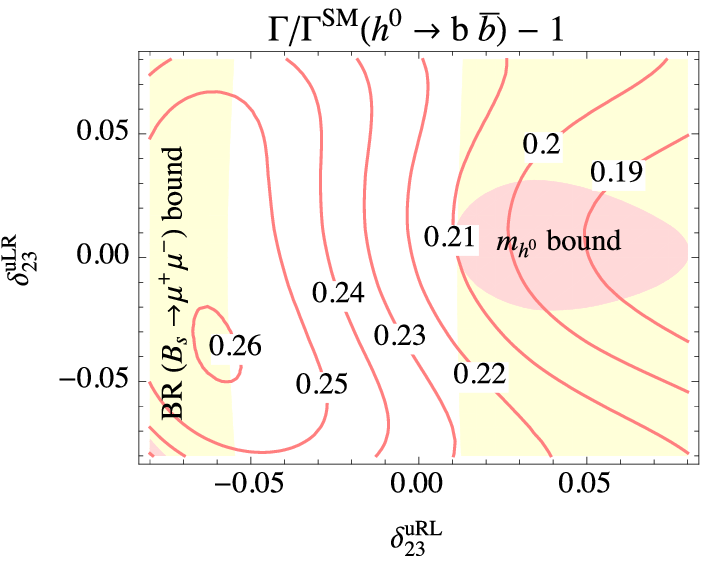}} \hspace*{0.8cm}}}
{ \mbox{\hspace*{-1cm} \resizebox{7.4cm}{!}{\includegraphics{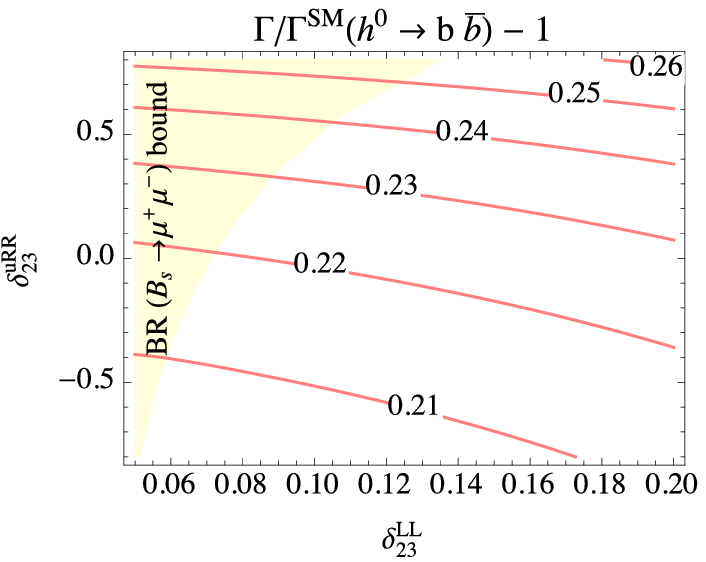}} }}
\caption{The relative difference of $\Gamma(h^0 \to b \bar{b})$ and $\Gamma^{\rm SM}(h^0 \to b \bar{b})$ as a function of the QFV parameters $\durl$ and $\dulr$ (left) and $\dll$ and $\durr$ (right). The shaded regions are excluded by the indicated constraints.
 }
\label{fig}
\end{figure*}

The dependence on the QFV parameters in the up-type squark sector is shown in Figure~\ref{fig}. On both plots the effect of QFV can go up to about 5\% mainly due to one-loop contributions with charginos. In particular, the dependence on the $\ti{c}_R-\ti{t}_L$ and the $\ti{c}_R-\ti{t}_R$ mixing parameters, $\durl$ and $\durr$, respectively, is stronger. 

We have also checked that for all $\d_{23}=0$ the deviation from the SM is already of about $20\%$ entirely due to QFC MSSM contributions. This has to be compared
to the case of $h^0 \to c \barc$~\cite{Bartl:2014bka}, where the MSSM QFC rate is compatible with the SM prediction and the deviation is mainly due to QFV. The latter is due to the different behaviour of the MSSM Higgs boson couplings to charm and to bottom quarks at large $\tan \beta$. 

The dependence on the QFV parameters in the down-type squark sector is negligible and therefore not shown here.

\section{Summary}

We have found that the full one-loop corrected width $\Gamma(h^0 \to b \barb)$ can differ up to about $26\%$ from its SM predicted value in a certain parameter region of the general MSSM. Up to $\sim 5\%$ of these are due to QFV loop contributions with large mixing between the second and the third generation up-squarks and large QFV trilinear couplings.
The rest of about $20\%$ difference is due to QFC MSSM contributions. In principle, such a difference can be measured e.g. at the ILC\cite{Tian:2013yda}. 

\section*{Acknowledgments}
This work is supported by the "Fonds zur F\"orderung der wissenschaftlichen Forschung (FWF)" of Austria, project No. P26338-N27.


\begin{thebibliography}{99}

\bibitem{HiggsAtlas}
G.~Aad et al. [ATLAS Collaboration], Phys.\ Lett.\ B {\bf 716} (2012) 1.

\bibitem{HiggsCMS}
S.~Chatrchyan et al. [CMS Collaboration], Phys.\ Lett.\ B {\bf 716} (2012) 30.

\bibitem{Bartl:2014bka}
  A.~Bartl, H.~Eberl, E.~Ginina, K.~Hidaka and W.~Majerotto,
  Phys.\ Rev.\ D {\bf 91} (2015) 1,  015007
  [arXiv:1411.2840 [hep-ph]].
  
%
  
 \bibitem{PDG} 
  K.~A.~Olive et al. (Particle Data Group),~Chin.\ Phys.\ C, {\bf 38}, 090001 (2014).
  
\bibitem{Casas:1996de}
  J.~A.~Casas and S.~Dimopoulos,
  Phys.\ Lett.\ B {\bf 387} (1996) 107
  [hep-ph/9606237].
  
\bibitem{Carena:1999py}
  M.~Carena, D.~Garcia, U.~Nierste and C.~E.~M.~Wagner,
  Nucl.\ Phys.\ B {\bf 577} (2000) 88
  [hep-ph/9912516].
  
\bibitem{Eberl:1999he}
  H.~Eberl, K.~Hidaka, S.~Kraml, W.~Majerotto and Y.~Yamada,
  Phys.\ Rev.\ D {\bf 62} (2000) 055006
  [hep-ph/9912463].
  
\bibitem{Tian:2013yda}
  J.~Tian {\it et al.} [ILD Collaboration],
  PoS EPS {\bf -HEP2013} (2013) 316
  [arXiv:1311.6528 [hep-ph]].
  
\end{thebibliography}
\end{document}